# Observation of multi-skyrmion objects created by size and density control in Ta/CoFeB/MgO films

Christian Denker[1], Sören Nielsen[2], Enno Lage[2], Malte Römer-Stumm[2], Hauke Heyen[1], Yannik Junk[1], Jakob Walowski[1], Konrad Waldorf[3], Markus Münzenberg[1], and Jeffrey McCord[2]

1. Institut für Physik, Universität Greifswald, Felix-Hausdorff-Straße 6, 17489 Greifswald, Germany

2. Christian-Albrechts-Universität zu Kiel, Institute for Materials Science, Nanoscale Magnetic Materials and Magnetic Domains, Kaiserstraße 2, 24143 Kiel, Germany

3. Institut für Mathematik und Informatik, Universität Greifswald, Walther-Rathenau-Str. 47, 17487 Greifswald, Germany

**Magnetic skyrmions are chiral spin textures with a nontrivial topology that offer a potential for future magnetic memory and storage devices. The controlled formation and adjustment of size and density of magnetic skyrmions in Ta/CoFeB/MgO trilayers is demonstrated. The material system is the ideal candidate for the use as a bottom electrode integration into CoFeB/MgO/CoFe magnetic tunnel junctions. By varying the CoFeB thickness close to the out-of-plane to in-plane magnetic phase transition, we find that subtle energy contributions enable the skyrmion formation in a narrow thickness window, corresponding to sub 10 pm variation in CoFeB thickness. Using magneto-optical imaging with quantitative image processing, variations in skyrmion diameter and distribution below the Abbe limit are analyzed. We demonstrate a high degree of skyrmion diameter and density control. Zero-field stable skyrmions can be set with proper magnetic field initialization. This demonstrated tunability and the degree of comprehension of skyrmion formation pave the way for future skyrmion based magnetic memory. Moreover, we demonstrate the controlled merging of individual skyrmions to complex topological objects. We compare our results with the baby-Skyrme model, developed to describe the soliton nature, for any topological charge n, and demonstrate the ability to form multi-skyrmion objects. These objects will be interesting for fundamental mathematical studies of the topological behavior of solitons in the future.**

Skyrme [1] predicted topologically stabilized wave objects to model nucleons in particle physics, which were termed skyrmions and exist in different media, including magnetic materials [2,3]. Magnetic skyrmions were first observed by using neutron diffraction in MnSi bulk material at cryostatic temperatures [4]. Besides materials with Dzyaloshinski-Moriya interaction (DMI) originating from bulk crystal symmetry breaking, likewise symmetry breaking by heavy metal/ferromagnetic (HM/FM) interfaces can generate the required DMI, thus allowing skyrmion formation in magnetic thin film systems with perpendicular magnetic anisotropy (PMA) [5]. There, the heavy metal and the ferromagnet determine the properties of skyrmions or skyrmion-like objects. In thin film systems, magnetic skyrmions exist as circular magnetic structures with a single spin pointing opposite to its carrying magnetic layer with a domain wall configuration leading to a non-vanishing winding number, also known as topological charge [6]. Zero-field skyrmions were first detected in Fe/Ir thin films by spin-polarized scanning tunneling microscopy [7]. In addition, they provided the first evidence for deterministic creation of individual skyrmions [8]. Magnetic films with PMA also form dipolar stabilized labyrinth domains but can exhibit topologically stabilized skyrmion bubbles with a fixed rotation sense with strong DMI or weaker DMI in combination with an applied magnetic field. Large skyrmions of several micrometers in diameter were generated at room temperature by the application of currents in patterned heavy metal/ferromagnet/oxide (HM/FM/Ox) trilayers [9] under the application of a magnetic bias field. Inspired by this breakthrough, skyrmion bubbles at room temperature and their generation have been investigated for many material systems. The material systems in which skyrmions could be established can be mainly classified in two groups: HM/FM/oxides with Co or CoFeB as the FM material [9,10,11,12,13] and $(HM/FM)_n$ multilayers with Co, Fe, Fe/Co as the repeated FM element [14,15,16,17,18]. For multilayer structures zero-field skyrmions were generated [19], but zero-field skyrmions could not be initialized for single FM layer structures. There, external magnetic bias fields for the stabilization of skyrmions are essential [20]. In addition to the application of external magnetic fields, exchange biased systems provide a similar stabilization mechanism [21]. Exceptions from this are limited to small patterned structures with local magneto-static skyrmion stabilization due to edge effects [11].

The intensive research activities on skyrmions are motivated by the quest for novel magnetic data storage solutions, with some resembling historic magnetic bubble technology [22,23]. Approaches are based on the shifting of magnetic domain walls by electrical currents in a resting host medium [24]. This drastically reduces the required size, energy consumption, and data access times. This idea was extended to the concept of a skyrmion racetrack memory [25]. Additional ideas on stochastic computing rely on the irreversible character of nucleation, movement, and annihilation of domain structures. In all cases magnetic textures must be generated, manipulated, and detected by electrical means. It has been shown that skyrmions can be created and shifted by current pulses [9,13,15]. The topological properties of skyrmions enable a variable toolbox for the detection, creation, and annihilation [26,27,28]. Skyrmions permit size reduction of the data carrying quasiparticles by orders of magnitude. Overall, they can be moved by electrical currents, which critical current density can be potentially five orders smaller as compared to domain wall racetrack memories [27].

Yet, for the envisioned practical applications, an electrical write- and read-out scheme is necessary. So far, the only realized electrical skyrmion detection is based on the topological Hall effect [29]. Alternatively, imaging techniques like spin polarized scanning tunneling microscopy at a low temperature around 4 K [8,30], electron microscopy (Lorentz transmission electron microscopy or scanning electron microscopy), synchrotron-based techniques using magnetic dichroism in the X-ray range, and magneto-optical Kerr effect (MOKE) imaging are used for the detection and investigation of skyrmion and skyrmion motion in various material systems (see review [3]). For practical applications spin transfer torque switching by magnetic tunnel junctions (MTJs) offers an efficient way for skyrmion creation and annihilation. In addition, MTJs offer a fast, scalable, and highly efficient option for skyrmion detection with excellent signal to noise ratio. MTJs for reading and writing of skyrmions have been proposed, but so far have not been experimentally realized [31]. Ta/CoFeB/MgO/CoFeB is a preferred base material system for MTJs with a theoretical maximum for the tunnel magnetoresistance (TMR) ratio of around 1000% [32]. The highest experimentally reached TMR ratio has been 602% for a system in-plane magnetic anisotropy (IMA) [33]. For our layer system we find 270% TMR ratio for in-plane and 60% for out-of-plane TMR devices. This makes Ta/CoFeB/MgO a highly interesting system for skyrmion research.

**SKYRMION GENERATION AND ZERO-FIELD SKYRMIONS**

In the following we demonstrate various ways of controlled skyrmion generation in the Ta/CoFeB/MgO MTJ bottom electrode material system. For that purpose, we avoided the paramagnetic (PM) to perpendicular magnetic anisotropy (PMA) transition for the ferromagnet thickness, where the formation of movable skyrmions is found because of paramagnetic instabilities at very low thickness [34], and where a reduced Curie temperature is exhibited. Instead, we investigated the parameter space for skyrmion generation at the out-of-plane (perpendicular anisotropy, PMA) to in-plane anisotropy (IMA) magnetic phase transition. This thickness regime has the critical advantage that the thermal stability at room temperature for device application is significantly increased. We explored the skyrmion size and skyrmion density modification with subtle changes in the ferromagnetic layer thickness and magnetic field variations, to seek for the suitability of MTJ based skyrmion writing and reading. In the following, we demonstrate that ferromagnetic film thickness control on the picometer thickness level is essential for reliable magnetic field induced and zero-field stable skyrmion generation. Varying the average thickness by only a few picometers, drastically influences skyrmion properties in a systematic way, and by this allows an unprecedented and high degree of control.

The presented studies lay the foundations for skyrmion manipulation in MTJ layer systems and their use in spintronic devices.

The dependence of skyrmion diameter with magnetic layer thickness and magnetic field relies on the subtle interplay of energies [35]. Close to the PMA to IMA magnetic phase transition, where effective PMA approaches zero, the effect of DMI is dominating the magnetic texture and skyrmions develop. Their size decreases while advancing to the magnetic phase boundary. Furthermore, the modeling suggests that with increasing magnetic out-of-plane field application the skyrmion size decreases until skyrmion annihilation at higher OOP magnetic bias fields. It becomes clear that the overall thickness range for the existence of skyrmions is very narrow with a high sensitivity of skyrmion properties to the magnetic film thickness. Not clear is how to nucleate skyrmions and how the transformation in skyrmion size takes place, also from a topological point of view. The nucleation of skyrmions is particularly of interest in conjunction with the alternative labyrinth domain state. Therefore, to generate and investigate skyrmion behavior, we perform our experiments in CoFeB layers with a linear FM thickness gradient, resulting in a corresponding magnetic anisotropy gradient. This allows to study effects arising in a small average thickness window of some pm of CoFeB thickness, as we demonstrate for the existence of the skyrmion phase. Especially, this enables us to study systematic variations of skyrmion properties without additional occurring parameter variations due to small sample-to-sample variations.

The schematics of the $Si/SiO_2/Ta/Co_{40}Fe_{40}B_{20}/MgO/Ru$ layer stack are shown in Fig. 1(a). Figure 1(b) displays a large field-of-view MOKE microscopy image of the sample in the demagnetized state for a film thickness range increasing from $t_{CoFeB} = 1.20$ nm to 1.44 nm (from the left to right). The linear FM thickness change over just 0.24 nm manifests itself in a gradual change from large extended magnetic domains millimeters in size to smaller sized domains. With increasing *FM thickness* in the PMA region the magnetic domain size shrinks from being clearly distinguishable, starting with some hundreds μm in size, down to below to the spatial resolution limit of the large view MOKE microscope. From the MOKE analysis the PMA to IMA transition is found to occur at $t_{CoFeB} = 1.384$ (23) nm. Above this critical thickness only in-plane (IP) magnetic domain contrast and no out-of-plane (OOP) MOKE contrast is visible, while strong OOP contrast and no IP contrast is observed below the CoFeB thickness of the PMA to IMA transition. The remanence magnetization $M_r/M_s$ and coercive field $H_c$ are nearly constant with $M_r \approx M_s$ with $\mu_0 H_c \approx 0.6 - 0.7$ mT for a CoFeB thickness up to 1.35 nm. The magnetic remanence decreases approaching the PMA to IMA transition. This is as well consistent with the formation of small magnetic domains, which become favorable with decreasing effective PMA with increasing CoFeB film thickness, approaching the IMA region. A corresponding linear MOKE signal amplitude change in the PMA regime clearly is in accordance with a linear thickness variation FM wedge structure (supplementary figure 1).

The magnetic domain structure evolution within this thickness regime is displayed in Fig. 1(c). Typical labyrinth or maze domains occur at zero magnetic fields (Fig. 1(c)-1). An analysis of the labyrinth domain width in this CoFeB film thickness range shows clearly the expected exponential decrease in domain width (supplementary figure 2) [36,37] with the decrease in domain wall energy, due to the PMA anisotropy strength in combination with an existing DMI. With the application of OOP magnetic fields, the labyrinth domain phase transforms into a mixed maze and skyrmion phase (Fig. 1(c)-2) and then into a pure skyrmion phase with higher OOP magnetic biasing fields (Fig. 1(c)-3). Generally, the transformation from a labyrinth to a skyrmion phase with the application of OOP field is a strong indication for the skyrmion nature of the developing magnetic texture. The skyrmion nature is further confirmed by their directional motion with applied current. The observation of skyrmions being stable under electric current induced motion. They show a large skyrmion Hall of up to 45° (supplementary figure 3). This indicates a non-vanishing winding number of the objects. For higher FM layer thickness in-plane domains with weaker MOKE contrast form (Fig. 1(c)-4), due to the predominant IMA.

In the PMA regime, the characteristics of the OOP magnetization curves display signatures of domain nucleation and annihilation. Magnetization loops are displayed in Fig. 1(d) (indicated in Fig. 1(b)). With increasing CoFeB thickness we find an increase of loop shearing and a corresponding decrease of hysteresis. The characteristic low PMA region in the narrow thickness range of $t_{CoFeB}$ from 1.361 nm to 1.383 nm is suitable for the nucleation and existence of narrow magnetic stripes and skyrmions during magnetization reversal. The existence of different magnetic domain phases with the application of magnetic fields during magnetization reversal is marked in the corresponding magnetization loops (compare also to Fig. 1(c)). Only labyrinth domains form for the lower thicknesses (Fig. 1(d)-1, $t_{CoFeB} = 1.361$ nm). With increasing $t_{CoFeB}$ (Fig. 1(d)-2, $t_{CoFeB} = 1.371$ nm), labyrinth domains nucleate from saturation. Yet, reversing the magnetic field, labyrinth and skyrmion domains coexist for applied field amplitudes above $\mu_0 H_z \approx 1.0$ mT. Close to saturation ($\mu_0 H_z \approx 2.2$ mT), even a pure skyrmion domain phase appears. With an only slight increase of the magnetic layer thickness by nominally just 6 pm the general magnetization loop characteristics remain (Fig. 1(d)-3, $t_{CoFeB} = 1.377$ nm). Yet, significant alterations in the resulting domain pattern characteristics are observed. Pronounced regions with labyrinth, coexisting labyrinth and skyrmion phases, as well as with a pure skyrmion phase, now appear for increasing and decreasing applied magnetic fields. Interestingly, the single domain and pure skyrmion phase are both stable in the field range around $\mu_0 H_z \approx 3$ mT, comparing the ascending and descending part of the magnetization loop. This magnetic history effect makes this range appealing for memory storage applications. A further increase of $t_{CoFeB}$ by just another 6 pm increases the magnetic field range of a skyrmion phase existence significantly (Fig. 1(d)-4, $t_{CoFeB} = 1.383$ nm). Concurrently, the magnetic field range for the appearance of the labyrinth domain phase is reduced to a narrow field regime. The manifestation of the mixed magnetic texture and the pure skyrmion phase is highly reproducible. Further increasing the CoFeB thickness results in an in-plane magnetization alignment due to an effective IMA (see also Fig. 1(c)-4) with IP domain alignment.

An overview of the manifestation of labyrinth and skyrmion phases with CoFeB thickness during magnetization reversal is summarized in the experimental phase diagram shown in Fig. 1(e), where different regimes of phase stability are indicated by color schemes. For OOP magnetic field applications three distinct regions ($I$ – $III$) are identified (compare to Fig. 1(c) and (d)). *Region I* represents the labyrinth phase and appears with small magnetic fields for all the regarded film thicknesses. This region overlaps with *region II*, showing the coexisting of labyrinth and skyrmions for a film thickness exceeding $t_{CoFeB}$ = 1.370 nm and magnetic fields above $\mu_0 H_z$ = 1 mT. This region partly overlaps with *region III*, which marks the parameter space for exclusive skyrmion formation, starting at film thicknesses above $t_{CoFeB}$ = 1.375 nm and magnetic fields exceeding about $\mu_0 H_z \approx 1.5$ mT in amplitude.

The skyrmion formation is dependent on magnetic history also in other ways. By combining OOP magnetic fields with IP magnetic field applications additional and different regions of stability, marked as A and B, are obtained. Combining the IP field application with small OOP magnetic fields result in a mixture of labyrinth domain and skyrmion phases (*region A*). Pure skyrmion phases are obtained for higher additional OOP magnetic field application (*region B*). The range of OOP magnetic field amplitude $H_z$ for the different regimes is indicated in Fig. 1(e). The range of magnetic field for *region B* is very narrow for 1.360 nm $\leq t_{CoFeB} \leq$ 1.375 nm for a magnetic bias field of around $\mu_0 H_z$ = 1.25 mT. With further increasing $t_{CoFeB}$ and an increase in the saturation field, the narrow band significantly widens up towards higher magnetic fields up to $\mu_0 H_z$ = 5 mT. This skyrmion phase exists until the sample exhibits IMA. None of the field applications lead to an entirely zero-field skyrmion phase. Similar magnetic history schemes for the prevention of labyrinth domain nucleation using IP magnetic field applications are known from PMA ferrite films [38]. This is the key to zero-field skyrmions. Their formation is facilitated by the use of pulsed IP fields. We show that by applying pulsed IP magnetic fields the nucleation and stabilization of a pure zero-field skyrmion phase (*region ZS*) in a narrow thickness range above and below $t_{CoFeB}$ = 1.375 nm is attained as further indicated in Fig. 1(e). However, so far, the existence of pure zero-field skyrmions in our single FM film structures is limited to a narrow $t_{CoFeB}$-window of roughly 10 pm CoFeB thickness variation. For larger thicknesses still in the PMA regime we obtain a zero-field mixed skyrmion and labyrinth phase (*regions ZM*).

In the following, we will discuss the dependence of skyrmion size and density as a function of $t_{CoFeB}$ and OOP magnetic field $H_z$. Intensity evaluation is used to determine the skyrmion diameter $d_{sky}$ of multiple skyrmions from the MOKE micrographs, allowing for a statistically relevant study of skyrmion behavior. The integral MO skyrmion intensity $I$ is equal to $I$ = $1/4 \cdot \pi \cdot d_{sky}^2 \cdot MO_\pm$, with $MO_\pm$ being the intensity difference between up and down magnetization normalized to its area. This is independent of the domain wall width, any blur by imaging optics or digital Gaussian noise filters because the equi-magnetization-line of 50% defines the skyrmion diameter under the assumption of symmetric domain walls. A standard segmentation algorithm with an intensity threshold to separate the individual domains and distinguish between the MOKE image regions belonging to a skyrmion and the surrounding magnetic matrix is used for automated skyrmion size and density analysis. Exemplary results from two skyrmions of different diameters are shown in Fig. 2. The threshold is chosen as close as possible to the average magnetic matrix MOKE intensity to include maximum pixels for the intensity sum for the derivation of skyrmion size, but high enough to allow for reliable separation of the magnetic matrix background from the image noise. By this scheme additional noise contributions increase the variation of the border of the detected skyrmion diameter, but generally retain the average diameter. The threshold of detection is kept at a minimal level to minimize the error of the diameter determination from neglecting pixels with minor MOKE intensity value. Accordingly, the noise level of the imaging system and the distance between the skyrmions define the minimum detectable diameter of skyrmions. The reliable free skyrmion diameter detection limit is around $d_{sky} \approx 100$ nm in our case, well below the spatial resolution of the MOKE microscope. The automatic image evaluation allows analysis of multiple skyrmions from a single magnetic domain image. In our case, up to hundreds of skyrmions per MOKE image could be analyzed and skyrmion densities and sizes are extracted for statistical analysis.

A comprehensive evaluation of the variation of types of domain textures, skyrmion density, and skyrmion diameter and distribution can be extracted from the analysis of large skyrmion ensembles. For a fixed CoFeB film thickness of $t_{CoFeB} = 1.379$ nm the application of magnetic fields is shown. The results of the analysis on the OOP magnetic field behavior are summarized in Fig. 3, top. Starting from a zero-field labyrinth domain structure, a mixture of labyrinth domains and skyrmions appear at an OOP field of about $\mu_0 H_z = 1$ mT (Fig. 3(a)-1). With further increasing the magnetic field, the density of labyrinth domains reduces in favor of skyrmions with a consequential increase of the skyrmion density to a maximum density of about $\rho_{sky} \approx 1$ $\mu m^{-2}$ at $\mu_0 H_z = 2$ mT (Fig. 3(b)). A stable pure skyrmion phase forms above that field (Fig. 3(a)-2, Fig. 3(a)-3), existing until magnetic saturation. Within that regime, increasing the OOP field results in a reduction of the skyrmion density ($\rho_{sky} = 0.06$ $\mu m^{-2}$ at $\mu_0 H_z = 3.7$ mT). Concurrently, the average skyrmion diameter decreases continuously from $d_{sky} \approx 400$ nm at $\mu_0 H_z = 1$ mT to $d_{sky} \approx 160$ nm (Fig. 3(c)). The skyrmion diameters are not constant but distributed (Fig. 3(d)) with a standard deviation from $d_{\sigma-sky} = 100$ nm down to $d_{\sigma-sky} = 50$ nm. Coinciding with the reduction in average skyrmion size, the extent of the skyrmion size distribution is decreasing.

What is the minimal attainable skyrmion size close to PMA to IP transition as we approach the CoFeB film thickness range from $t_{CoFeB} \approx 1.361$ nm to 1.382 nm we can create and detect? The skyrmions were initialized by an IP field pulse application under an OOP magnetic bias field.

We show in Fig. 3, bottom, the size and density dependence as a function of CoFeB thickness along the boundary between skyrmion and saturated magnetization state in the phase stability diagram. Accordingly, the OOP magnetic field is increased to the maximum for each position with the ability to observe skyrmions along this stability line. Exemplary MOKE images showing isolated skyrmions are displayed in Fig. 3(e). The variance in the skyrmion density shown in Fig. 3(f) is strongly dependent on how close the OOP magnetic field is relative to the critical field for skyrmion annihilation. The average skyrmion diameter $d_{sky}$, shown in Fig. 3(g), decreases with increasing $t_{CoFeB}$ from $d_{sky} = 750$ nm at $t_{CoFeB} = 1.361$ nm and $\mu_0 H_z = 0.8$ mT to $d_{sky} = 135$ nm at $t_{CoFeB} = 1.382$ nm and $\mu_0 H_z = 5$ mT. This is accompanied by a decrease of the skyrmion diameter variation from about $d_{\sigma-sky} = 210$ nm down to $d_{\sigma-sky} = 40$ nm. The extracted size distribution shows a clear cut off at the resolution limit of our skyrmion identification algorithm at $d_{sky} = 85$ nm for the highest $t_{CoFeB}$ and accordingly smallest skyrmion diameter. Yet, the extracted size distribution (Fig. 3(h)) indicates the existence of even smaller skyrmions, the diameter value of which is not accessible by our systematic analysis procedure.

For device applications stable skyrmions at zero magnetic bias field and at room temperature are a necessary precondition. In continuous PMA films and in the examples analyzed above the magnetic textures are usually introduced through an OOP variation of the magnetic field. Stabilization and enabling skyrmions without external magnetic fields have been only achieved with the assistance of geometrical confinement in patterned nanostructures [11] due to magnetostatic interactions. In such films skyrmions have been so far only observed in the presence of an OOP magnetic bias field.

## MULTI SKYRMION STATES

This high degree of control of skyrmion size and density in our wedged sample system, with picking the skyrmion state with picometer thickness resolution, allows us to use it as a skyrmion playground to force their interaction and to form a multi-skyrmion state. A multi-skyrmion state can be controllably formed by starting from skyrmions, each with topological charge $n = 1$, and merging them by interaction to multi-skyrmion solitons. Due to the topological protection, the total topological charge of the merged topological soliton is expected to be the sum of the individual topological charges. This is backed up by a mathematical analysis of the spherical topology of the system and has moreover been tested numerically by Weidig [39] in a so-called baby-Skyrme model. We have selected a numerical example for five solitons, each of charge $n = 1$, relaxing to multi-skyrmions with higher charges. In that example, one finds two multi-skyrmions with $n = 2$ and one with $n = 1$ on the right (Fig. 4(a)). To realize this in a real sample, one needs a high degree of control of the skyrmion interaction. We have chosen a regime, where the generation of zero-field skyrmions is achieved by a magnetic field applied nearly along the magnetic hard axis and nearly parallel to the sample surface. After the application of the slightly tilted IP fields, magnetic skyrmions are accomplished. The development of skyrmions with reducing the magnetic field to the remanent skyrmion state is shown in Fig. 4(b). The application of a $H_{1°} \approx 100$ mT nearly IP magnetic field (tilted OOP by 1°) results in skyrmions for $t_{CoFeB}$ ranging from 1.363 nm to 1.374 nm as also indicated in Fig. 1(e) (stars, *region ZS*). The corresponding OOP field component in which skyrmions are observed is around $H_z = 2$ mT within the range of skyrmion existence with pure OOP field application. However, reducing the applied field to zero preserves the initial skyrmion phase (Fig. 4(b)-1 to -3). Instead of forming labyrinth domains, the decrease of the applied magnetic field is accompanied by an increase in skyrmion size together with a decrease in skyrmion density. While the skyrmion density decreases from $\rho_{sky} \approx 0.40$ µm$^{-2}$ to $\rho_{sky} \approx 0.16$ µm$^2$ (Fig. 4(c)), the average diameter increases from $d_{sky} = 0.34$ µm to $d_{sky} = 1.35$ µm (Fig. 4(d)). This forces the coalescence of skyrmions. For better illustration, traced skyrmions are color coded in Fig. 4(b). We assume for all skyrmions the same starting charge, since we know that all show the same direction of the skyrmion Hall angle close to 45° (supplementary figure 3). Like in the numerical prediction we find a multi-skyrmion state. Some skyrmions stay with a topological charge, one merges to a charge of larger than $n = 2$, and one shows the merging of three skyrmions, consequently to a topological charge larger than $n = 3$. This indicates homogenous growth and is expected for random coalescence. We find that the thermal diffusion is low and the lifetimes of the (metastable) skyrmions in the range of several hours to days.

**CONCLUSIONS**

The stability of magnetic skyrmions as well as their size and density control in Ta/CoFeB/MgO trilayers mimicking the bottom electrode and tunnel barrier of an MTJ are demonstrated. Using a CoFeB wedge structure allows precise and continuous magnetic anisotropy tuning. By this we were able to stabilize skyrmions close to the PMA to IMA transition. Skyrmions appear as part of OOP magnetic field variation at a small $t_{CoFeB} = 1.370 - 1.383$ nm window from nearly zero OOP fields up to around $\mu_0 H_z \approx 4.5$ mT. Skyrmion generation by in-plane field pulses extends the parameter range for skyrmions to thicker CoFeB layers as well as to an extended magnetic field range. For both approaches a phase diagram is derived. The skyrmion diameter and density can be tuned by the effective anisotropy defined by the CoFeB film thickness and external bias field. The average diameter is adjustable from below 100 nm skyrmion diameter up to about 1.4 µm, while the density can be freely tuned from below 1 per µm² to single skyrmion density. The merging of skyrmions is qualitatively compared to the baby-Skyrme model.

Zero-field stable skyrmions in single structures are stabilized by a setting routine comprising the application of nearly in-plane magnetic fields, encompassing skyrmion coalescence. The possibility of allowing for controllable adjustment of single domain and skyrmion states at zero fields together with the high stability of the states, the long skyrmion lifetimes, and appropriate pinning, demonstrates the general suitability of this material system and the tunability of skyrmion properties for magnetic memory applications.

## Methods

**Thin film preparation.** Samples with a variation of FM CoFeB thickness $t_{CoFeB}$ were grown as wedges with nominally picometer thickness control. The films were grown in a UHV cluster system under the same optimized conditions as for the fabrication of Ta/CoFeB/MgO MTJs [40] on thermally oxidized Si(100) substrates. A sample of Si(100)/ 500 nm $SiO_2$/ 5-3.7 nm (wedge) Ta/ 1.20-1.56 nm (wedge) $Co_{40}Fe_{40}B_{20}$/ 3.2nm MgO/ 3 nm Ru is investigated (Fig. 1(a)). The Ta and CoFeB layers are prepared by magnetron sputtering deposition using growth rates of 0.078 nm/s and 0.045 nm/s, respectively, in Ar atmosphere at a pressure of $5 \cdot 10^{-3}$ mbar and a base pressure around $5 \cdot 10^{-10}$ mbar. To obtain a gradual thickness variation, the samples are mounted under an angle of 50° with respect to the sputter source normal. This geometric factor is resulting in a 20% linear decrease for the Ta thickness, and most important a 20% linear increase of the CoFeB layer thickness along the substrate. The resulting thickness gradient for a CoFeB film with a center thickness of 1.2 nm is 0.028 nm/mm. For the Ta film with a thickness of 5 nm in the center, this results in an opposite thickness gradient of -0.116 nm/mm across the substrate. The MgO and Ru layers are deposited in-situ by e-beam evaporation in the UHV cluster with a base pressure in the range of $10^{-9}$ mbar and a deposition rate of 0.02 nm/s. The Ru layer prevents degradation of MgO at atmosphere. Ex-situ annealing is performed in vacuum with a base pressure below $10^{-7}$ mbar for 1 h at final temperature. During the process of sample optimization, we tested several variations in the sample preparation and post-annealing. The sample of investigation was annealed for 1 h at 60 °C to initialize the solid-state-epitaxy process at the MgO/CoFeB interface as for the MTJs.

**Magneto-optical characterization.** Magneto-optical Kerr effect (MOKE) based magnetometry was used to extract local magnetization loops. MOKE imaging [41] is performed by a large-view MOKE microscope for further pre-characterization of the magnetic properties and for the exact identification of the position of the PMA to IMA transition on the wedged layers samples. High-resolution MOKE microscopy with variable and combined IP and OOP magnetic field application is used for the imaging of the skyrmion characteristics.

**Magnetic texture analysis.** From the MOKE micrographs, magnetic texture image analysis is performed using ImageJ 1.51g including the Fiji image processing packages [42]. For the quantitative analysis of skyrmion density and size the MOKE micrographs are segmented with a standard threshold algorithm and evaluated by the standard Fiji function for particle analysis. The MOKE contrast is related to the magnetization $m = M/M_s$, which is obtained by normalizing to the maximum MOKE contrast derived from the local MOKE intensity difference between up and down magnetized magnetic domains, corresponding to the magnetization amplitude $\pm M_s$.

Taking the normalization into account, the intensity threshold that corresponds to the magnetic skyrmion matrix, is manually adjusted. The effective skyrmion diameter is determined by the integral of MOKE intensity of the skyrmion particle subtracted by the background intensity of the surrounding magnetic matrix. The method is similar to the method used for the determination of integral domain wall widths in soft magnetic samples [43] and skyrmion diameters in HM/FM multilayers [44].

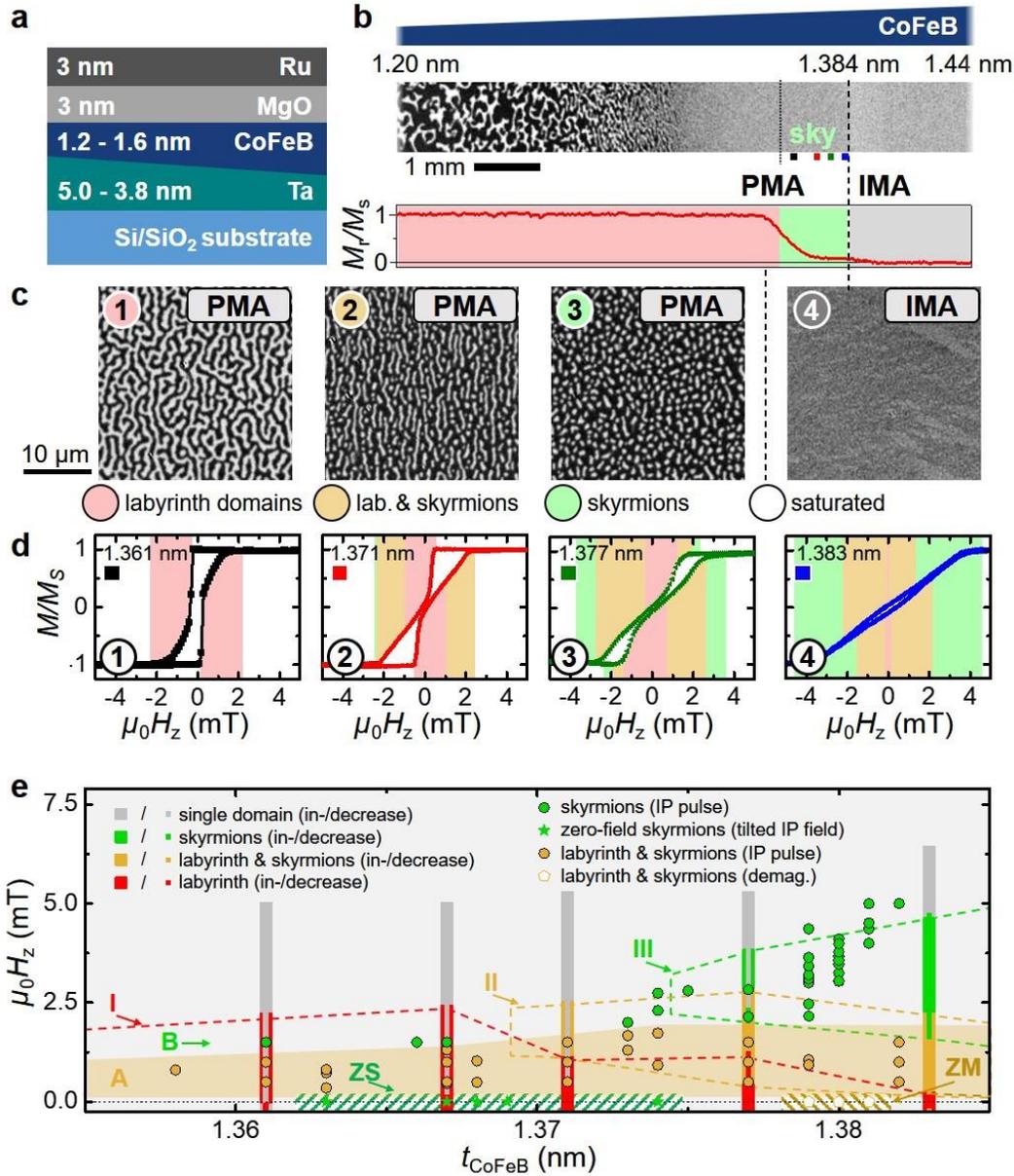

**Figure 1. Magnetic domain evolution and phase diagram at the perpendicular magnetic anisotropy region (PMA) to the in plane magnetic anisotropy region (IMA).** (a) Sketch of the Si(100)/ 500 nm SiO$_2$/ 5-3.7 nm (wedge) Ta/ 1.20-1.56 nm (wedge) Co$_{40}$Fe$_{40}$B$_{20}$/ 3.2 nm MgO/ 3 nm Ru layer stack. (b) Overview MOKE microscopy image of the demagnetized state across the CoFeB wedge. The CoFeB thickness and the perpendicular magnetic anisotropy region (PMA), including the region able to host skyrmions (sky), and the in plane magnetic anisotropy region (IMA) are indicated. (c) Exemplary magnetic domain states including labyrinth domains (lab), mixed labyrinth and skyrmions (lab & sky), and skyrmion (sky) phases (as indicated). (d) OOP magnetization loops at selected CoFeB thickness positions. The positions along the CoFeB wedge are accordingly color coded in (b). The marked colored regions indicate the observed domain phases (red: labyrinth, yellow: labyrinth and skyrmions,

green: skyrmions). (e) Phase diagram. **I** (green dashed line): Skyrmions appearing directly by OOP field variation, **II** (orange dashed line): Labyrinth domains and skyrmions forming directly by OOP field variation, **III** (red dashed line): Labyrinth domains from OOP field variation. The colored areas indicate the parameter space for skyrmions generated by IP magnetic pulses with a concurrent OOP bias field: **A** (orange): Labyrinth domains and skyrmions, **B** (green): Pure skyrmion phase. Zero-field skyrmion stabilization. **ZM** (striped orange): Mixed labyrinth domains and skyrmions at zero magnetic field, **ZS** (striped green): Stable pure skyrmion domain states at zero field. Zero-field skyrmions are induced by decreasing a slightly tilted amplitude 100 mT IP to zero.

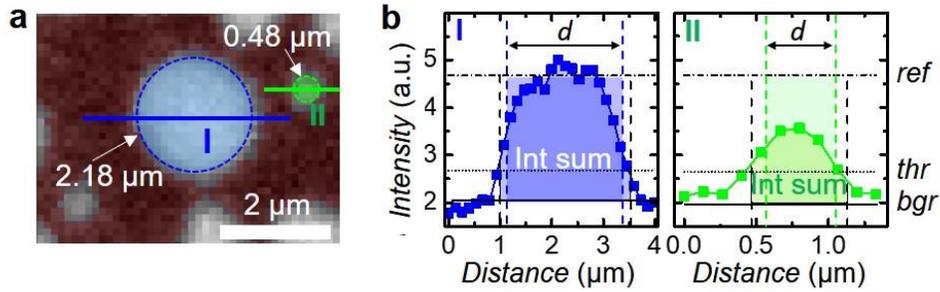

**Figure 2. Image analysis by threshold segmentation and size determination by integrated intensity beyond the Abbe limit.** (a) MOKE image with pixels below the threshold shaded dark reddish. The determined diameters for two exemplary skyrmions are indicated by shaded circles with dashed outlines in blue ($d_{sky}$ = 2.18 µm) and green ($d_{sky}$ = 0.48 µm). (b) Corresponding line intensity plots as indicated in (a). The background intensity (*bgr*), corresponding to the magnetic matrix MOKE intensity, and the threshold intensity (*thr*) applied for the automated analysis are indicated. The average maximum intensity for a fully opposite magnetization to the matrix is shown as the shaded area below the curves to visualize the integrated intensity. The shaded rectangles display the determined skyrmion diameter, taking the difference from the measured intensity maximum to the maximum (*ref*) signal into account.

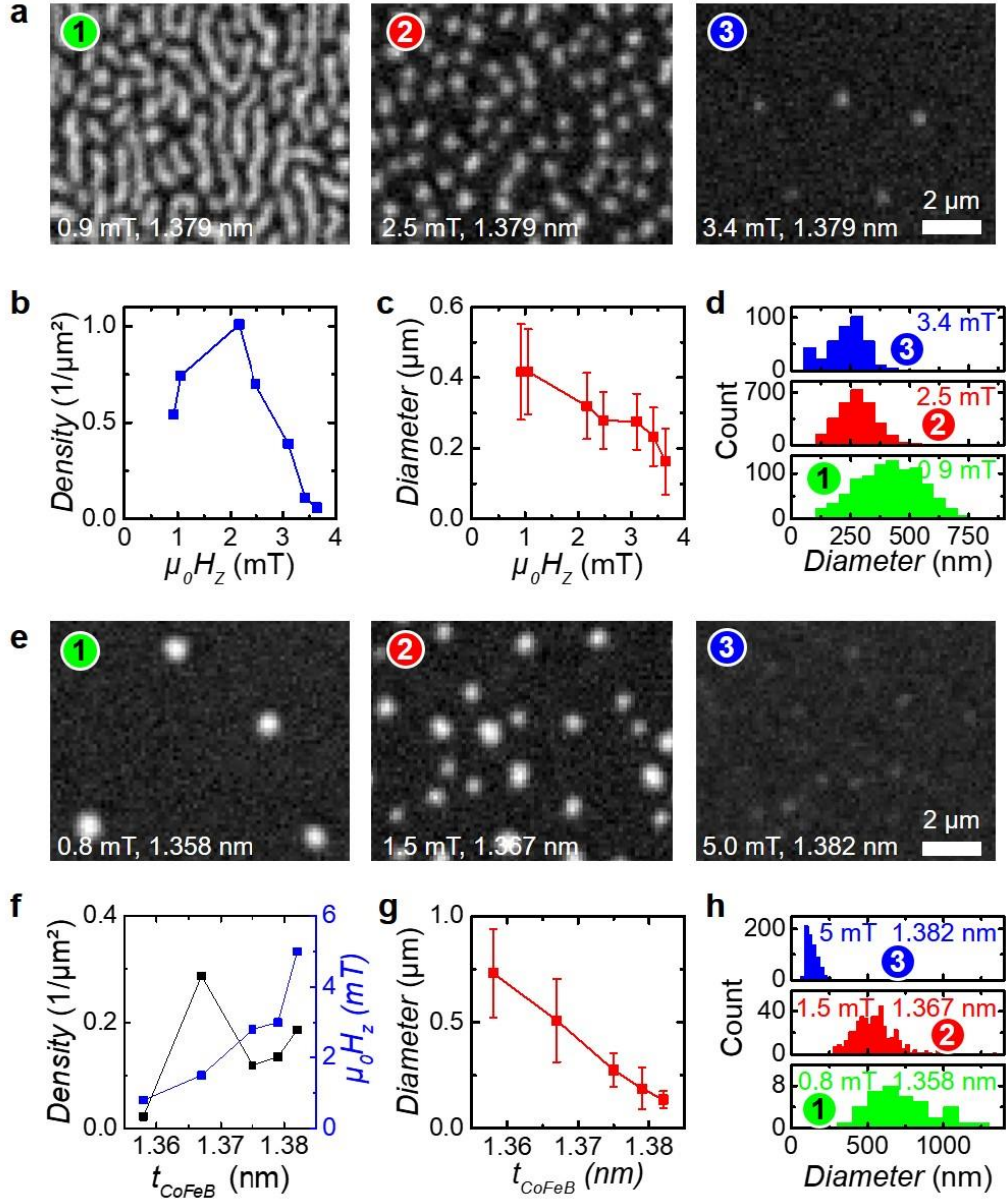

**Figure 3. Large skyrmions ensemble analysis: magnetic field and thickness.** *Magnetic domains induced by the application of IP magnetic field pulses as function of the OOP bias field at a CoFeB thickness of 1.379 nm.* (a) MOKE microscopy images for an OOP magnetic bias field of $\mu_0 H_z$ = 0.9 mT, 2.5 mT, and 3.4 mT as indicated. The brightness and contrast were adjusted to cover the range from the minimum background intensity to the maximum bright intensity. (b) Skyrmion density and (c) diameter as a function of the applied OOP magnetic bias field. The error bars show the standard deviation of the obtained skyrmion diameter distribution. (d) Corresponding histograms of the skyrmion diameter distributions.

*Magnetic domains induced by application of IP magnetic field pulses as function of the CoFeB thickness with the OOP magnetic field close to instability respectively the spatial resolution limit of imaging.* (e) MOKE microscopy images at a CoFeB thickness of 1.379 nm for OOP

bias field of $\mu_0 H_z$ = 0.8 mT, 1.5 mT, and 5.0 mT. (f) Skyrmion density and corresponding OOP magnetic field amplitude as a function of CoFeB thickness. (g) Skyrmion diameter changes with CoFeB thickness. The error bars show the standard deviation of the skyrmion diameter distribution. (h) Corresponding histograms of the skyrmion diameter distributions.

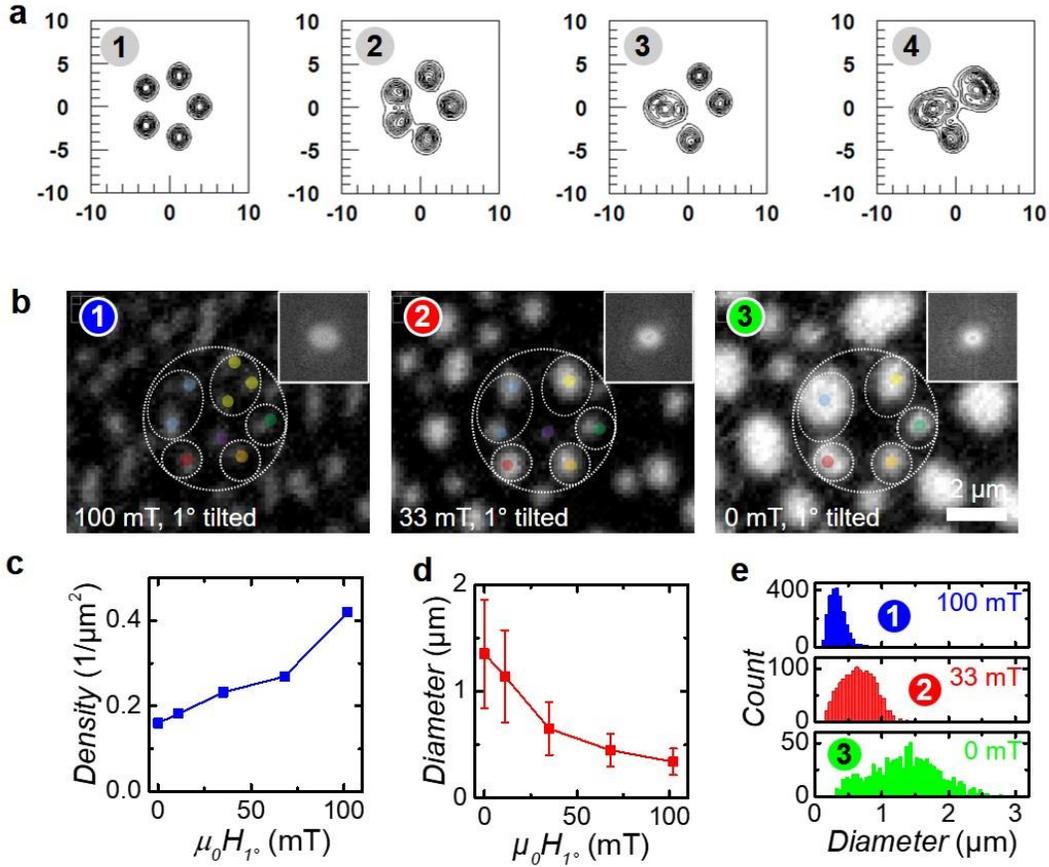

**Figure 4. Multi-skyrmions of higher topological charge.** *Creation of multi-skyrmions in the baby-Skyrme model.* (a) Multi-skyrmion objects evolution in the baby-Skyrme model as predicted by Weidig numerically for five solitons of charge $n = 1$, relaxing to multi-skyrmions with charge $n > 1$ (two with $n = 2$ and one of $n = 1$ [35]). A merging of skyrmions is observed. *Magnetic skyrmions development by decreasing a 100 mT by 1° tilted IP magnetic field down to zero.* (b) MOKE microscopy images at a CoFeB thickness of 1.363 nm at three states of magnetic field application. Magnetic field values are indicated. MOKE image brightness and contrast are adjusted for better skyrmion visibility. The colored spots and dashed circles mark groups of skyrmions that coalesce during the procedure. (c) and (d) show the skyrmion density and diameter as a function of the applied magnetic field. The error bars in (d) correspond to the standard deviation of skyrmion diameter distributions, which are depicted in (e).

## ACKNOWLEDGMENTS

The project was initially financially supported by the "Bund Norddeutscher Universitäten" and later funded by the German Research Foundation (DFG) through the Priority Program SPP 2137 Skyrmionics: "Topological Spin Phenomena in Real-Space for Applications". J.M. and M.M. thank K. Brandenburg for careful proof-reading of the manuscript.

## AUTHOR CONTRIBUTION STATEMENT

M.M., K.W., and J.M. initiated and designed the research. C.D., H.H. and J.W. deposited the multiplayer. C.D., Y.J. and S.N. conducted the magnetic property characterization. S.N., M.R.-S., Y.J, E.L. and J.M. carried out the MOKE microscopy. C.D. and Y.J. wrote the code for and analyzed the skyrmion size distributions. J.M., K.W. and M.M. supervised the project. C.D., J.M. and M.M. wrote the manuscript. All authors discussed the results and reviewed the manuscript.

**Supplementary information**

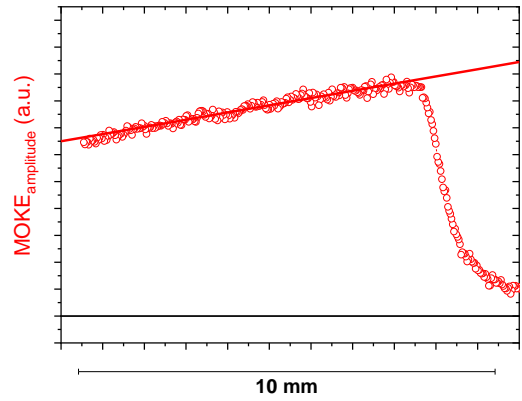

**Figure S1.** Relative and linear change of saturation magnetization MOKE contrast (MOKE$_{amplitude}$) as a function of sample position parallel to the CoFeB thickness gradient.

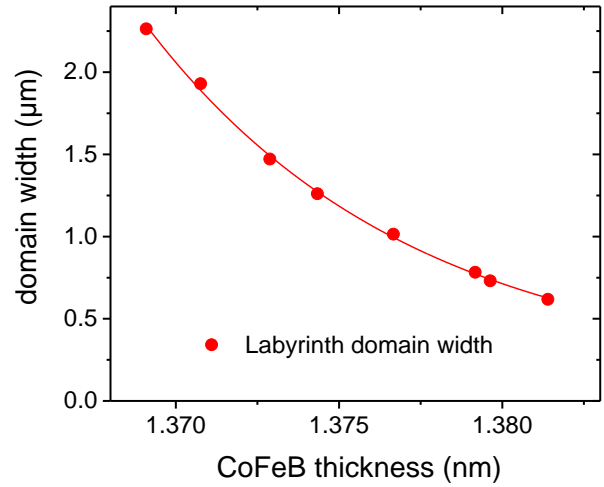

**Figure S2.** Zero field labyrinth domain width versus CoFeB thickness, displaying an exponential decrease in domain width with ferromagnetic layer thickness. Note that the skyrmion size is considerably smaller.

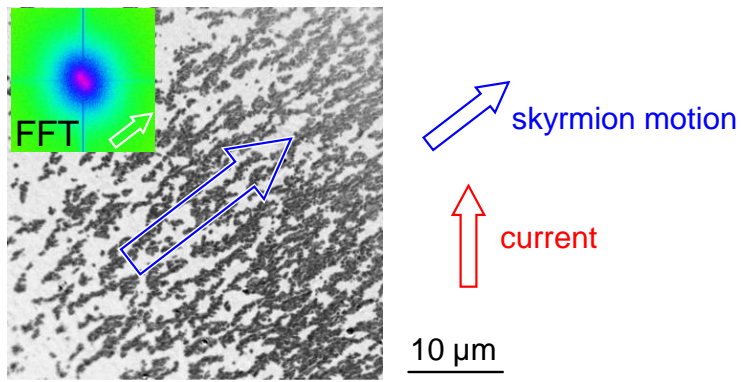

**Figure S3.** Recorded skyrmion tracks, displaying the overall skyrmion motion from a series of 70 current pulses (upward current pulses, pulse width 50 ns, current density approx. $10^{11}$ A/cm$^2$). The direction of average skyrmion motion is indicated.